\documentclass[a4paper,12pt]{article}
\usepackage{amsmath}
\usepackage{amssymb}
\usepackage{amsthm}
\usepackage{mathtools}
\usepackage{alphabeta}
\usepackage[newzealand]{babel}
\usepackage{booktabs}
\usepackage[margin=15pt,font={footnotesize},labelfont=bf,format=hang]{caption}
\usepackage[T1]{fontenc}
\usepackage[hmargin=1in,top=0.8in,bottom=1in]{geometry}
\usepackage{graphicx}
\usepackage[hidelinks]{hyperref}
\usepackage{cleveref}
\usepackage{microtype}
\usepackage{natbib}
\usepackage{wasysym}
\usepackage{xcolor}

\newcommand{\uppi}{\text{\pi}}

\newcommand{\kron}{\delta}

\makeatletter
    \newcommand\rcvector[2][\\]{\ensuremath{%
      \global\def\rc@delim{#1}%
        \negthinspace\begin{pmatrix}
          \rc@vector #2,\relax\noexpand\@eolst%
        \end{pmatrix}}}
    \def\rc@vector #1,#2\@eolst{%
      \ifx\relax#2\relax
        #1
      \else
        #1\rc@delim
        \rc@vector #2\@eolst%
      \fi}
\makeatother
\renewcommand{\binom}{\rcvector}

\makeatletter
    \renewcommand{\section}{\@startsection
    {section}
    {1}
    {0mm}
    {-\baselineskip}
    {0.5\baselineskip}
    {\large\bfseries}} 
\makeatother

\begin{document}	
\title{Generating, from scratch, the near-field asymptotic forms of scalar resistance functions for two unequal rigid spheres in low Reynolds number flow}
\author{Adam K.\ Townsend\thanks{adam.k.townsend@durham.ac.uk}\\ \\Department of Mathematical Sciences, Durham University,\\ Stockton Road, Durham DH1 3LE, United Kingdom}
\maketitle
\begin{abstract}
The motion of rigid spherical particles suspended in a low Reynolds number fluid can be related to the forces, torques and stresslets acting upon them by 22 scalar resistance functions, commonly notated $X^A_{11}$, $X^A_{12}$, $Y^A_{11}$, etc. Near-field asymptotic forms of these resistance functions were derived in Jeffrey and Onishi (J.\ Fluid Mech., 1984) and Jeffrey (Phys.\ Fluids A, 1992); these forms are now used in several numerical methods for suspension mechanics. However, the first of these important papers contains a number of small errors which make it difficult for the reader to correctly evaluate the functions for parameters not explicitly tabulated. This short article comprehensively corrects these errors, and adds formulae that were originally omitted from both papers, so that the reader can verify and implement the equations independently. The corrected expressions, rationalised and using contemporary nondimensionalisation, are shown to match mid-field values of these scalars which are calculated through an alternative method. A Python script to generate and evaluate these functions is provided.
\end{abstract}

\section{Introduction}
The linearity of Stokes flow allows the forces, torques and stresslets acting on particles suspended in a low Reynolds number fluid to be related to the particles' velocities, angular velocities and fluid background strain rate through a resistance or mobility matrix. For rigid spherical particles, the elements of the resistance matrix can be formed from scalar resistance functions, commonly notated as
\begin{equation}
\begin{split}
&X^A_{11}, X^A_{12}, Y^A_{11}, Y^A_{12}, Y^B_{11}, Y^B_{12}, X^C_{11}, X^C_{12}, Y^C_{11}, Y^C_{12},\\
&X^G_{11}, X^G_{12}, Y^G_{11}, Y^G_{12}, Y^H_{11}, Y^H_{12}, X^M_{11}, X^M_{12}, Y^M_{11}, Y^M_{12}, Z^M_{11}, Z^M_{12}. \end{split}\label{scalars}
\end{equation}
The near-field asymptotic forms of the scalar resistance functions for two unequal rigid spheres were initially published in \citet{jeffrey_calculation_1984} and \citet{jeffrey_calculation_1992}. The terms in the asymptotic forms themselves rely on recurrence relations, and, in the first of these two papers, there are a number of small errors. Furthermore, the reader may find it difficult to generate the required expressions (and therefore the value of these functions) in both papers independently, given the omission of intermediate formulae.

Some of these errors appear to have been noticed by authors, including indirectly by \citet{kim_microhydrodynamics:_2005} in corrected tabulations, and a partial list of errata has been published online by \citet{ichiki_errata_2008}. However, a comprehensive, fully corrected description of how to generate expressions for these functions, to enable verifiable calculation for arbitrary size ratios and separation distances, is yet to be published.

This short article is a compilation of the relevant equations, with those originally omitted now added, and with any errors fixed. Equations from \citet{jeffrey_calculation_1984} are labelled as (JO 1.1); similarly, equations from \citet{jeffrey_calculation_1992} are labelled as (J 1), and those from the helpful \citet{ichiki_resistance_2013} are labelled as (I 1). Pages from \citet{kim_microhydrodynamics:_2005} are labelled as (K\&K, p.\,1).

At the end of the article, the questions of error identification and validation of the corrected forms are addressed.

Throughout this article, we use the same notation as the original papers. For two spheres of radius $a_1$, $a_2$ with centres a distance $s$ apart, we define the non-dimensional gap $\xi$ and size ratio $\lambda$ as
\begin{equation}
	\xi = \frac{2s}{a_1 + a_2} - 2, \qquad \lambda = \frac{a_2}{a_1}.
\end{equation}
The near-field forms are valid for $\xi \ll 1$ and $\xi \ll \lambda$.

The nondimensionalisation in (JO 1.7) and (J 3) scales the resistance functions by factors of $(a_1+a_2)/2$. As noted in \citet{jeffrey_calculation_1992}, although this choice provides symmetry elsewhere, it leads to unwieldy factors of $(1+\lambda)^{-1}$ in some of the expressions we are interested in. Here, we instead follow the convention of \citet[\S 11.3]{kim_microhydrodynamics:_2005}, and scale all functions by factors of $a_1$. Specifically, to recover the dimensional forms, it is necessary to multiply the scalars in \cref{scalars}, derived as follows, by an appropriate factor:
\begin{equation}
\begin{tabular}{lcccccc}
\toprule
\text{superscript in the scalar name} & $A$ & $B$ & $C$ & $G$ & $H$ & $M$ \\
\midrule
\text{multiply by} & \,$6 \uppi a_1$\, & \,$4 \uppi a_1^2$\, & \,$8 \uppi a_1^3$\, & \,$4 \uppi a_1^2$\, & \,$8 \uppi a_1^3$\, & \,$\displaystyle\frac{20}{3}\uppi a_1^3$\,\\
\bottomrule
\end{tabular}
\tag{3, 4}
\end{equation}
\setcounter{equation}{4}

A Python script implementing the corrected formulae is available on GitHub \citep{github}.

\section{$\textit{X}^\textit{\,A}$ terms}
Here the $X^A$ formulae are given in full, with changes from the source material when noted. The same directions for alteration, when required, will be given for the other terms in later sections.

We first set up up the recurrence relations from (JO 3.6--3.9),
\begin{align}
	P_{n00} &= \delta_{1n},\\
	V_{n00} &= \delta_{1n},\\
	V_{npq} &= P_{npq} - \frac{2n}{(n+1)(2n+3)}\sum_{s=1}^q \binom{n+s,n} P_{s(q-s)(p-n-1)},\\
	P_{npq} &= \sum_{s=1}^q \binom{n+s,n} \left(\frac{n (2 n+1) (2 n s-n-s+2)}{2 (n+1) (2 s-1) (n+s)}P_{s(q-s)(p-n+1)} \right. \nonumber\\
    	&\hspace{17ex}{}-\frac{n (2 n-1)}{2 (n+1)}P_{s(q-s)(p-n-1)} \\
	&\hspace{16.4ex}\left.{}-\frac{n (4 n^2-1)}{2 (n+1) (2 s+1)}V_{s(q - s - 2) (p - n + 1)} \right). \nonumber
\end{align}
Then, we define the formulae from (JO 3.15 and 3.19),
\begin{align}
	f_k(\lambda) &= 2^k \sum_{q=0}^k P_{1(k-q)q}\lambda^q,\label{f-k-xa}\\
	g_1(\lambda) &= 2\lambda^2(1+\lambda)^{-3},\\
	g_2(\lambda) &= \frac{1}{5}\lambda(1+7\lambda+\lambda^2)(1+\lambda)^{-3},\\
	g_3(\lambda) &= \frac{1}{42}(1+18\lambda - 29\lambda^2 + 18\lambda^3+\lambda^4)(1+\lambda)^{-3}, \\
	m_1(m) &= -2\kron_{m2} + (m-2)(1-\kron_{m2}),\label{m1-function}
\end{align}
We then define two intermediate functions, the first of which is given in (JO 3.22) as
\begin{align}
	A^X_{11} &= 1 - \frac{1}{4}g_1 + \sum_{\mathclap{\substack{m=2\\ m\text{ even}}}}^\infty \left[ 2^{-m}(1+\lambda)^{-m}f_{m}-g_1-2m^{-1}g_2 + 4m^{-1}m_1^{-1}g_3 \right]. \label{ax11}
\end{align}
The function $m_1$ is somewhat awkward as, despite being a toggle for whether $m=2$, it is used in \citet{jeffrey_calculation_1984} for summation over both even and odd values of $m$. The later paper, \citet{jeffrey_calculation_1992}, removes it by using instead a form of the summation which here would render the equivalent expression
\begin{equation}
	A^X_{11} = 1 - \frac{1}{4}g_1 - g_3 + \sum_{\mathclap{\substack{m=2\\ m\text{ even}}}}^\infty \left[ 2^{-m}(1+\lambda)^{-m}f_{m}-g_1-2m^{-1}g_2 + 4m^{-1}(m+2)^{-1}g_3 \right].
	\tag{14a}
	\label{removem1}
\end{equation}
It is computationally convenient to have the same coefficients of $f_m$, $g_1$, $g_2$ and $g_3$ (where they exist) in the summations for all functions of this type, so we choose to use this form when describing corrections going forward. Consistent coefficients also allow us to notice that all $f_m$ functions are defined with a factor of $2^m$ which cancels the $2^{-m}$ in the $f_m$ coefficient in the sum: these powers of 2 can therefore be left out of numerical implementations. In practice, we also find that moving the factor $(1+\lambda)^{-m}$ inside the sum in the definition of $f_m$ can prevent the terms in the sum (specifically from $\lambda^q$) from exceeding the maximum float size when $\lambda$ and $m$ (and hence $q$) are large. However, we do neither of these things in this paper to remain consistent with the original texts.

The second intermediate function is from (JO~3.23), and indeed needs correcting to be
\begin{align}
	-\frac{1}{2}(1+\lambda) A^X_{12} &= \frac{1}{4}g_1 + 2g_2\log2-2g_3\nonumber\\
	& \quad + \sum_{\mathclap{\substack{m=1\\m\text{ odd}}}}^\infty \left[ 2^{-m}(1+\lambda)^{-m}f_m - g_1 - 2m^{-1}g_2 + 4m^{-1}(m+2)^{-1}g_3 \right],
\end{align}
noting the change from $m_1$ (which is identically $m-2$ given that $m$ is odd) to $m+2$. 

Then the resistance scalars are given by
\begin{align}
	X_{11}^A &= g_1 \xi^{-1} + g_2 \log (\xi^{-1}) + A^X_{11} + g_3 \xi \log(\xi^{-1}),\\
	-X_{12}^A &= g_1 \xi^{-1} + g_2 \log(\xi^{-1}) - \frac{1}{2}(1+\lambda)A^X_{12} + g_3 \xi \log(\xi^{-1}),
\end{align}
from (JO 3.17--3.18) up to $O(\xi\log(\xi^{-1}))$: note the different scaling on $X^A_{12}$ as we use the aforementioned convention from (K\&K, p.\,279).
\section{$\textit{Y}^\textit{\,A}$ terms}
The recurrence relations are (JO 4.6--4.11), but with $V_{npq}$ corrected to
\begin{equation}
	V_{npq} = P_{npq} + \frac{2n}{(n+1)(2n+3)}\sum_{s=1}^q \binom{n+s,n+1} P_{s(q-s)(p-n-1)},
\end{equation}
noticing the sign change on the $1$ in the last subscript.

The required intermediate formulae for the $f$, $g$ and $m$ functions are \cref{f-k-xa}, (JO 4.16--4.17), and \cref{m1-function}, respectively.

Then the $A^Y$ terms are given by (JO 4.17--4.18): if desired, these can be more conveniently written to remove $m_1$ and use the \citet{jeffrey_calculation_1992} form by writing
\begin{align}
	A^Y_{11} &= 1 - g_3 + \sum_{\mathclap{\substack{m=2\\ m\text{ even}}}}^\infty \left[ 2^{-m}(1+\lambda)^{-m}f_{m}-2m^{-1}g_2 + 4m^{-1}(m+2)^{-1}g_3 \right],
	\tag{18a} \\
	-\frac{1}{2}(1+\lambda) A^Y_{12} &= 2g_2\log2 - 2g_3\nonumber\\
	& \quad + \sum_{\mathclap{\substack{m=1\\m\text{ odd}}}}^\infty \left[ 2^{-m}(1+\lambda)^{-m}f_m - 2m^{-1}g_2 + 4m^{-1}(m+2)^{-1}g_3 \right], 
	\tag{18b}
\end{align}
but this does not constitute a correction.

Either way, we are left with the resistance scalar formulae,
\begin{align}
	Y^A_{11} &= g_2 \log(\xi^{-1}) + A^Y_{11} + g_3 \xi \log (\xi^{-1}),\\
	-Y^A_{12}&= g_2 \log(\xi^{-1}) - \frac{1}{2}(1+\lambda)A^Y_{12} + g_3 \xi \log (\xi^{-1}),
\end{align}
from (JO 4.15--4.16), with a different scaling on $Y^A_{12}$.

\section{$\textit{Y}^\textit{\,B}$ terms}
The recurrence relations are the same as those for the $Y^A$ terms. The required intermediate formulae for the $f$ and $g$ functions are
\begin{equation}
	f_k(\lambda) = 2^{k+1}\sum_{q=0}^k Q_{1(k-q)q} \lambda^q,
\end{equation}
and (JO between 5.6 and 5.7), respectively. 

The $B^Y$ terms are corrected from (JO 5.7--5.8) to become
\begin{align}
	B^Y_{11} &= 2g_2\log2 - 2g_3 \nonumber \\
	& \quad + \sum_{\mathclap{\substack{m=1\\ m\text{ odd}}}}^\infty \left[ 2^{-m}(1+\lambda)^{-m}f_{m}-2m^{-1}g_2 + 4m^{-1}(m+2)^{-1}g_3 \right],\\
	-\frac{1}{4}(1+\lambda)^2 B^Y_{12} &= -g_3\ + \sum_{\mathclap{\substack{m=2\\m\text{ even}}}}^\infty \left[ 2^{-m}(1+\lambda)^{-m}f_m - 2m^{-1}g_2 + 4m^{-1}(m+2)^{-1}g_3 \right].
\end{align}

Then the resistance scalars are given by
\begin{align}
	Y^B_{11} &= g_2 \log(\xi^{-1}) + B^Y_{11} + g_3 \xi \log (\xi^{-1}),\\
	-Y^B_{12} &= g_2 \log(\xi^{-1}) - \frac{1}{4}(1+\lambda)^2B^Y_{12} + g_3 \xi \log(\xi^{-1}),
\end{align}
from (JO 5.5--5.6), with a different scaling on $Y^B_{12}$.
\section{$\textit{X}^\textit{\,C}$ terms}
Expressions for the resistance scalars can be expressed directly as
\begin{align}
	X^C_{11} &= \frac{\lambda^3}{(1+\lambda)^3}\zeta\left(3,\frac{\lambda}{1+\lambda}\right)-\frac{\lambda^2}{4(1+\lambda)}\xi\log(\xi^{-1}),\\
	X^C_{12} &= -\frac{\lambda^3}{(1+\lambda)^3}\zeta(3,1)+\frac{\lambda^2}{4(1+\lambda)}\xi\log(\xi^{-1}),\label{rohilla-fix}
\end{align}
from (JO 6.9--6.10), where $X^C_{12}$ uses the different scaling, and where $\zeta(z,a)$ is the Hurwitz zeta function,
\begin{equation}
	\zeta(z,a) = \sum_{k=0}^\infty \frac{1}{(k+a)^z}.
\end{equation}
\section{$\textit{Y}^\textit{\,C}$ terms}
The recurrence relations are the same as those for the $Y^A$ terms except that the initial conditions are replaced by (JO 7.3--7.5). 

The intermediate formula for the $f$ function is
\begin{align}
	f_k(\lambda) &= 2^k \sum_{q=0}^k Q_{1(k-q)q}\lambda^{q + (k \text{ mod } 2)},
\end{align}
with the $g$ formula given by (JO between 7.10 and 7.11), with the correction to $g_5$ of
\begin{align}
	g_5(\lambda) &= \frac{2}{125}\lambda(43-24\lambda+43\lambda^2)(1+\lambda)^{-4}.
\end{align}

Then the $C^Y$ terms from (JO 7.11--7.12), correcting the latter [note the power of $(1+\lambda)$] and writing both to avoid $m_1$, are
\begin{align}
	C^Y_{11} &= 1 - g_3 + \sum_{\mathclap{\substack{m=2\\ m\text{ even}}}}^\infty \left[ 2^{-m}(1+\lambda)^{-m}f_{m}-2m^{-1}g_2 + 4m^{-1}(m+2)^{-1}g_3 \right],\\
	C^Y_{12} &= 2g_4\log2 - 2g_5\ + \sum_{\mathclap{\substack{m=1\\m\text{ odd}}}}^\infty \left[ 2^{3-m}(1+\lambda)^{-3-m}f_m - 2m^{-1}g_4 + 4m^{-1}(m+2)^{-1}g_5 \right]. \label{koch-fix}
\end{align}

The resistance scalars are finally
\begin{align}
	Y^C_{11} &= g_2 \log(\xi^{-1}) + C^Y_{11} + g_3\xi\log(\xi^{-1}),\\
	\frac{8}{(1+\lambda)^3}Y^C_{12} &= g_4 \log(\xi^{-1}) + C^Y_{12} + g_5 \xi \log(\xi^{-1}),\label{y12c}
\end{align}
again with a different scaling on $Y^C_{12}$ compared to (JO 7.9--7.10).

To write \cref{koch-fix} in the \citet{jeffrey_calculation_1992} summation form, introduce $\tilde{g}_4$ and $\tilde{g}_5$ such that $\tilde{g}_{4} = (1+\lambda)^{3}g_{4}/8$ (similarly for $\tilde{g}_5$). This changes \cref{y12c} to become consistent with the other scalar definitions,
\begin{align}
	Y^C_{12} &= \tilde g_4 \log(\xi^{-1}) + \frac{1}{8}(1+\lambda)^3 C^Y_{12} + \tilde g_5 \xi \log(\xi^{-1}),
	\tag{34a}
\end{align}
which is the choice made in (K\&K, p.\,283). Note that the aforementioned error in $g_5$ persists here: the `$250$' in the denominator of the $\xi \log (\xi^{-1})$ term in $Y_{12}^C$ (what we are calling $\tilde{g}_5$) should read `$500$'.
\section{$\textit{X}^\textit{\,G}$ terms}
The recurrence relations are the same as those for $X^A$, and the $f$ and $g$ functions are (I 94) and (J between 19b and 20a). The $G^X$ terms are given by (J 21), noting that in their notation, $\tilde f(\lambda) = 2^{-m}f(\lambda)$. This gives us expressions for $X^G$ of
\begin{align}
	X^G_{11} &= g_1 \xi^{-1} + g_2 \log(\xi^{-1})+ G^X_{11} + g_3 \xi \log (\xi^{-1}),\\
	X^G_{12}&= -g_1 \xi^{-1} - g_2 \log(\xi^{-1}) + \frac{1}{4}(1+\lambda)^2 G^X_{12} - g_3 \xi \log(\xi^{-1}),
\end{align}
from (J 19), with a different scaling on the $X^G_{12}$.
\section{$\textit{Y}^\textit{\,G}$ terms}
The recurrence relations are the same as those for $Y^A$, and the $f$ and $g$ functions are (I 115) and (J between 27b and 28a). The $G^Y$ terms are given by (J 29), giving us expressions for $Y^G$ of
\begin{align}
	Y^G_{11} &= g_2 \log(\xi^{-1})+ G^Y_{11} + g_3 \xi \log (\xi^{-1}),\\
	Y^G_{12}&=  -g_2 \log(\xi^{-1}) + \frac{1}{4}(1+\lambda)^2 G^Y_{12} - g_3 \xi \log(\xi^{-1}),
\end{align}
from (J 27), with a different scaling on the $Y^G_{12}$.
\section{$\textit{Y}^\textit{\,H}$ terms}
The recurrence relations are the same as those for $Y^C$, and the $f$ and $g$ functions are (I 120) and (J between 35b and 36a). The $H^Y$ terms are given by (J 37), giving us expressions for $Y^H$ of
\begin{align}
	Y^H_{11} &= g_2 \log(\xi^{-1})+ H^Y_{11} + g_3 \xi \log (\xi^{-1}),\\
	Y^H_{12}&=  g_5 \log(\xi^{-1}) + \frac{1}{8}(1+\lambda)^3 H^Y_{12} + g_6 \xi \log(\xi^{-1}),
\end{align}
from (J 35), with a different scaling on the $Y^H_{12}$.
\section{$\textit{X}^\textit{\,M}$ terms}
The recurrence relations are the same as those for $X^A$, but with the different initial conditions (J 44). The $f$ and $g$ functions are given by (I 105) and (J between 48b and 49a). The $M^X$ terms are given by (J 50), giving us expressions for $X^M$ of
\begin{align}
	X^M_{11} &= g_1 \xi^{-1} + g_2 \log(\xi^{-1}) + M^X_{11} + g_3 \xi \log(\xi^{-1}),\\
	X^M_{12} &= g_4 \xi^{-1} + g_5 \log(\xi^{-1}) + \frac{1}{8}(1+\lambda)^3 M^X_{12} + g_6 \xi \log(\xi^{-1}),
\end{align}
from (J 48), with a different scaling on the $X^M_{12}$.
\section{$\textit{Y}^\textit{\,M}$ terms}
The recurrence relations are the same as those for $Y^A$, but with the different initial conditions (J 58). The $f$ and $g$ functions are given by (I 125) and (J between 64b and 65a). The $M^Y$ terms are given by (J 66), giving us expressions for $Y^M$ of
\begin{align}
	Y^M_{11} &= g_2 \log(\xi^{-1}) + M^Y_{11} + g_3 \xi \log(\xi^{-1}),\\
	Y^M_{12} &= g_5 \log(\xi^{-1}) + \frac{1}{8}(1+\lambda)^3 M^Y_{12} + g_6 \xi \log(\xi^{-1}),
\end{align}
from (J 64), with a different scaling on the $Y^M_{12}$.
\section{$\textit{Z}^\textit{\,M}$ terms}
The recurrence relations are (J 73--76). The $f$ and $g$ functions are given by (I 131) and (J between 79b and 80a). The $M^Z$ terms are given by (J 81), giving us expressions for $Z^M$ of
\begin{align}
	Z^M_{11} &= M^Z_{11} + g_3 \xi \log(\xi^{-1}),\\
	Z^M_{12} &= \frac{1}{8}(1+\lambda)^3 M^Z_{12} - g_3 \xi \log(\xi^{-1}),
\end{align}
from (J 79), with a different scaling on the $Z^M_{12}$.
\section{Error identification}
The original articles provide tabulated values of the intermediate scalars $A^X_{11}$, etc. The easiest way for the reader to confirm mistakes in the formulae is to confirm that the values computed from these formulae do not match those tabulated. Consider the following example:
\begin{center}
\begin{tabular}{lccc}
\toprule
 & \; value from (JO) formulae \; & \; value in (JO) tables \; & \; correct value \; \\
 \midrule
 $A^X_{12}(\lambda=1)$ & $-0.243\,00$ & $-0.350\,22$ & $-0.350\,22$ \\[0.8mm]
 $B^Y_{11}(\lambda=1)$ & $-0.8355\hphantom{\,0}$ & $-0.2390\hphantom{\,0}$ & $-0.2390\hphantom{0\,}$ \\
\bottomrule
\end{tabular}
\end{center}
Happily, the errors in the formulae for the intermediate scalars appear not to extend greatly to the tabulation of their values for a small range of size ratios in \citet[\S 11.3]{kim_microhydrodynamics:_2005}, where we find agreement with values calculated from the expressions in this article to at least the second significant figure (and often more). 

Nonetheless, errors in the formulae can also be spotted by either deriving the equations independently or observing that the values derived do not match those in the mid-field; both methods are described as follows.

\section{Validation of corrected expressions}
The reader is invited to derive and confirm the above formulae themselves, should they wish. The method is perhaps best explained in \citet{jeffrey_calculation_1992}, \S\S II (starting at the paragraph containing the definition of $\xi$), III\,B and III\,C, where $X^G$ is used as an example.

\begin{figure}
\centering
  Resistance scalars for $\lambda = 1$
  
  \vspace{1mm}
  \includegraphics[width=1.00\textwidth]{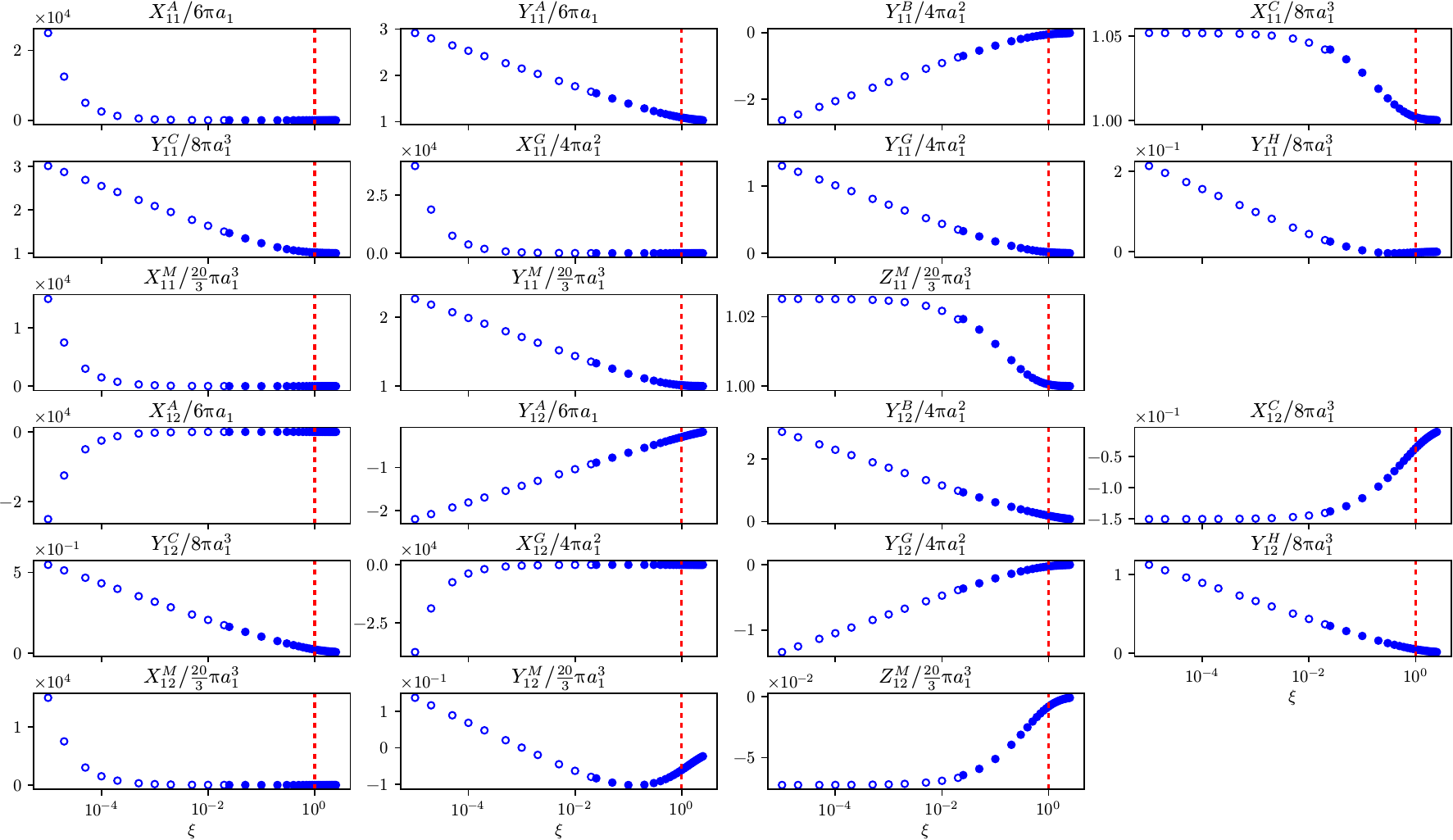}
  \caption{Values of the scalar resistance functions over non-dimensional gap, $\xi$, for size ratio $\lambda=1$. Those generated from the near-field formulae are represented by hollow circles ({\color{blue}\Circle}), and those generated from Lamb's solution \citep{wilson_stokes_2013} are filled circles ({\color{blue}\CIRCLE}). The dashed vertical line appears at $\xi=\lambda$, recalling that the near-field formulae are only valid for $\xi \ll \lambda$.}
  \label{fig:xyz-lam1}
\end{figure}

\begin{figure}
\centering
  Resistance scalars for $\lambda = 0.1$
  
  \vspace{1mm}
  \includegraphics[width=1.00\textwidth]{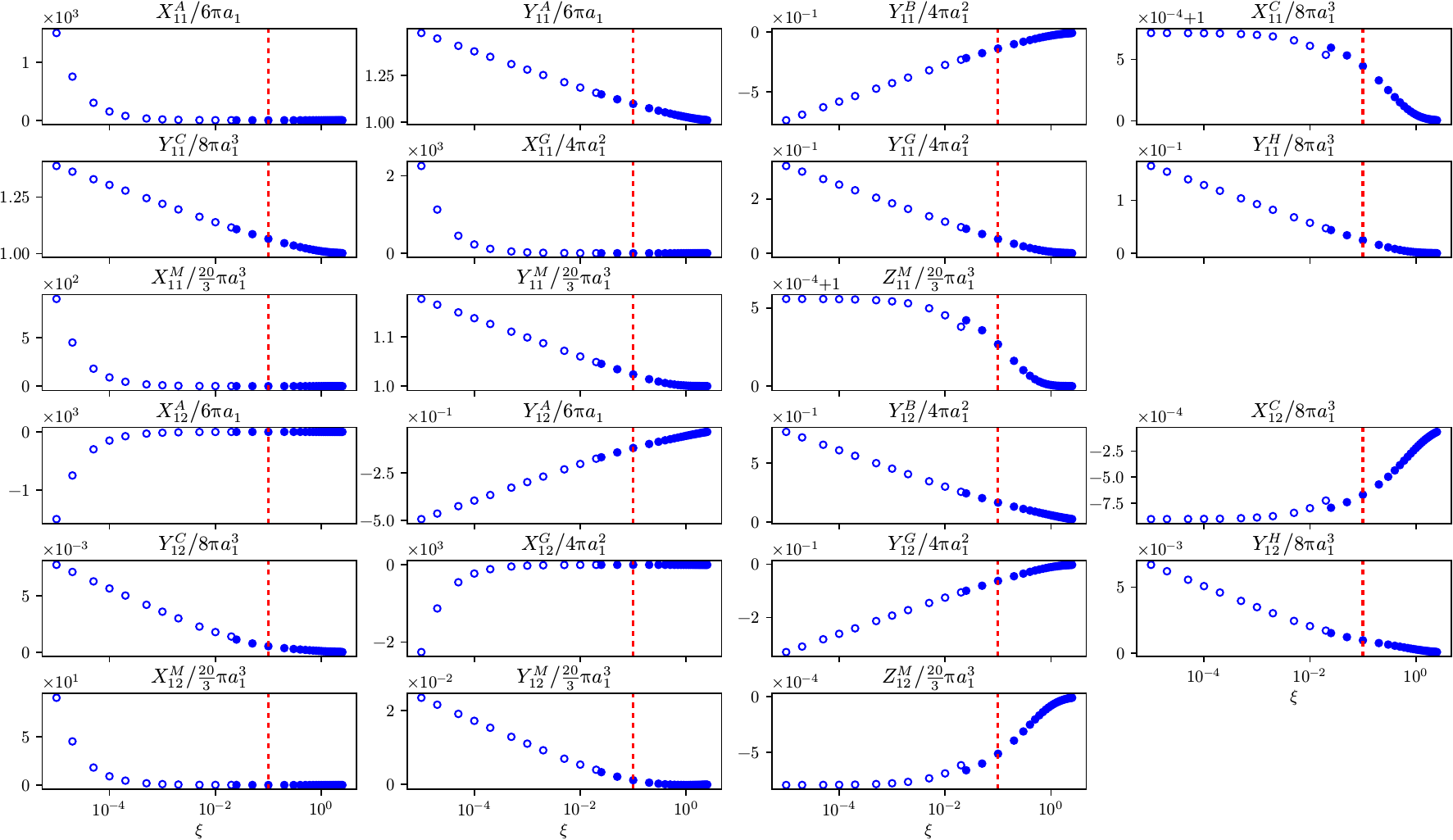}
  \caption{Values of the scalar resistance functions over non-dimensional gap, $\xi$, for size ratio $\lambda=0.1$. Those generated from the near-field formulae are represented by hollow circles ({\color{blue}\Circle}), and those generated from Lamb's solution \citep{wilson_stokes_2013} are filled circles ({\color{blue}\CIRCLE}). The dashed vertical line appears at $\xi=\lambda$, recalling that the near-field formulae are only valid for $\xi \ll \lambda$.}
  \label{fig:xyz-lam0p1}
\end{figure}

\begin{figure}
\centering
  Resistance scalars for $\lambda = 0.01$
  
  \vspace{1mm}
  \includegraphics[width=1.00\textwidth]{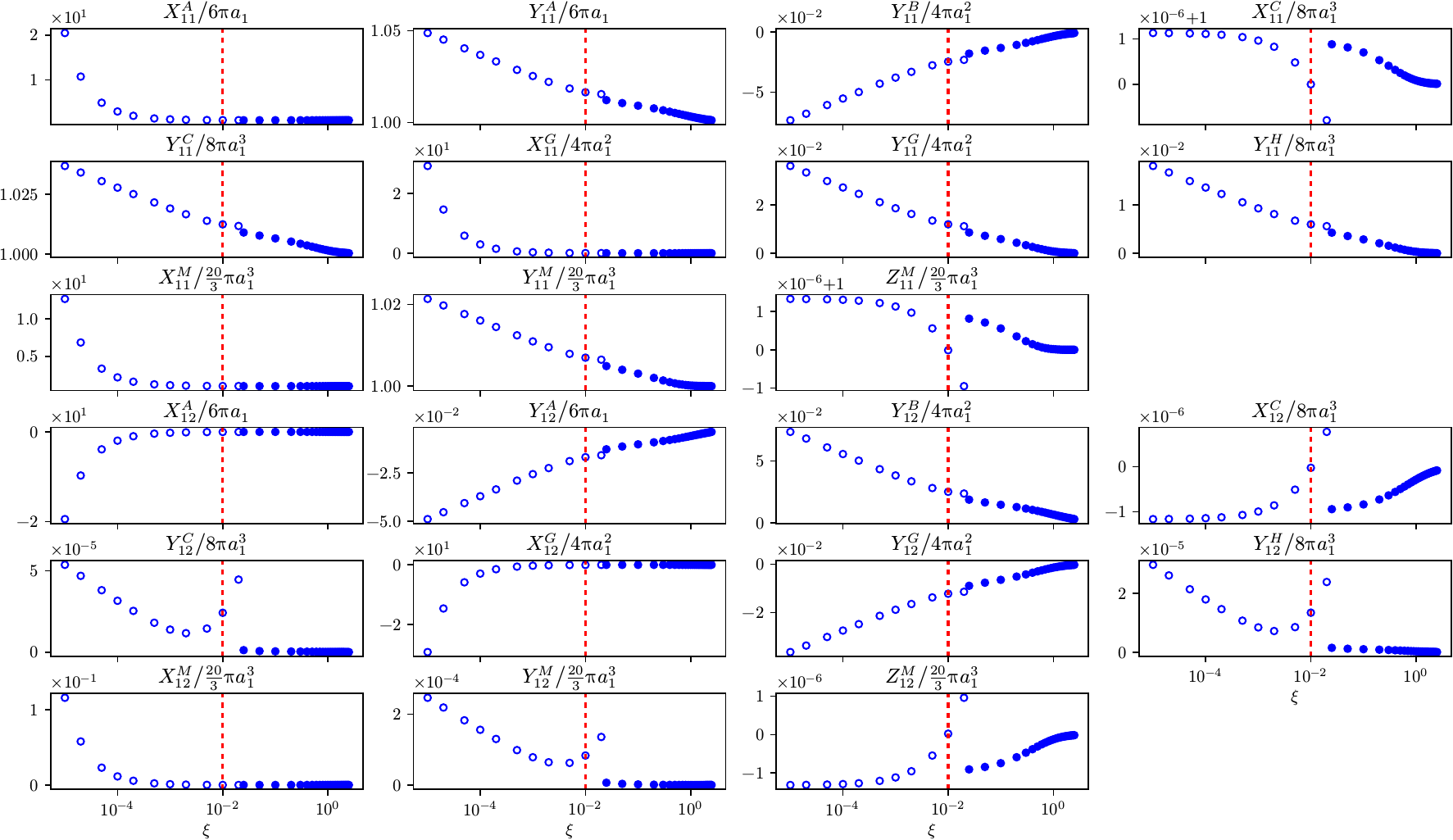}
  \caption{Values of the scalar resistance functions over non-dimensional gap, $\xi$, for size ratio $\lambda=0.01$. Those generated from the near-field formulae are represented by hollow circles ({\color{blue}\Circle}), and those generated from Lamb's solution \citep{wilson_stokes_2013} are filled circles ({\color{blue}\CIRCLE}). The dashed vertical line appears at $\xi=\lambda$, recalling that the near-field formulae are only valid for $\xi \ll \lambda$. Discrepancies at the meeting point of the near-field and mid-field values are consistent with the near-field series expansion ignoring terms of $O(\xi)$.}
  \label{fig:xyz-lam0p01}
\end{figure}

We can also confirm that the formulae produce values of the resistance scalars which match those in the mid-field. \Cref{fig:xyz-lam1,fig:xyz-lam0p1,fig:xyz-lam0p01} demonstrate the near-field values matching to the mid-field values, which have been computed independently for $\xi \gtrsim 0.014$ using the two-sphere method of \citet{wilson_stokes_2013}, based on the solution to Stokes flow given by \citet{lamb_hydrodynamics_1932}. In computing the near-field values, infinite sums in the expressions (e.g., in \cref{ax11}) are truncated after 100 nonzero terms; the addition of further terms does not change the appearance of the graphs. Recall that the near-field series expansions are valid only for $\xi \ll \lambda$ and ignore terms of $O(\xi)$.

\section*{Acknowledgments}
The author is thankful to Anubhab Roy (IIT Madras), Donald L.\ Koch (Cornell), Pankaj Rohilla (Texas Tech), Yixiang Luo and Aleksandar Donev (NYU) for correspondence regarding earlier drafts of the manuscript.

\small
\bibliography{references}{}
\bibliographystyle{jfm2}


\end{document}